\documentstyle[12pt]{article}
\newcommand{\bce}{\begin{center}}
\newcommand{\ece}{\end{center}}
\newcommand{\beq}{\begin{equation}}
\newcommand{\eeq}{\end{equation}}
\newcommand{\bea}{\vspace{0.25cm}\begin{eqnarray}}
\newcommand{\eea}{\end{eqnarray}}

\newcommand{\ba}{\begin{array}}
\newcommand{\ea}{\end{array}}

\newcommand{\doublespace}{
    \renewcommand{\baselinestretch}{1.6}\large\normalsize}

\def\lsim{\mathrel{\rlap{\lower4pt\hbox{\hskip1pt$\sim$}}
    \raise1pt\hbox{$<$}}}     
\def\gsim{\mathrel{\rlap{\lower4pt\hbox{\hskip1pt$\sim$}}
    \raise1pt\hbox{$>$}}}     

\def\lsim{\mathrel{\rlap{\lower4pt\hbox{\hskip1pt$\sim$}}
    \raise1pt\hbox{$<$}}}         
\def\gsim{\mathrel{\rlap{\lower4pt\hbox{\hskip1pt$\sim$}}
    \raise1pt\hbox{$>$}}}         

\def\beq{\begin{equation}}
\def\endeq{\end{equation}}
\def\arr{\begin{eqnarray}}
\def\endarr{\end{eqnarray}}

\makeindex
\begin{document}
\phantom{.}\hspace{8.cm}{\Large \bf KFA-IKP(Th)-1995-02} \\
\phantom{.}\hspace{9.5cm}{\large \bf  January 1995}
\vspace{2cm}
\begin{center}
{\bf \huge Quadrupole deformation of deuterons and final state
interaction in $^2\vec{H}(e,e'p)$ scattering on tensor
polarized deuterons at CEBAF energies\\}
\vspace{1cm}
{\bf A.Bianconi$^{1,2)}$, S.Jeschonnek$^{3)}$,
N.N.Nikolaev$^{3,4)}$, B.G.Zakharov$^{3,4)}$ } \medskip\\
{\small \sl
$^{1)}$Istituto Nazionale di Fisica Nucleare,
Sezione di Pavia, Pavia, Italy \\
$^{2)}$Dipartimento Fisica Nucleare e Teorica,
Universit\`a di Pavia, Italy \\
$^{3)}$IKP(Theorie), Forschungszentrum  J\"ulich GmbH.,\\
D-52425 J\"ulich, Germany \\
$^{4)}$L.D.Landau Institute for Theoretical Physics, \\
GSP-1, 117940, ul.Kosygina 2, V-334 Moscow, Russia
\vspace{1cm}\\}
{\bf \LARGE A b s t r a c t \bigskip\\}
\end{center}
The strength of final state interaction (FSI) between
struck proton and spectator neutron in $^2\vec{H}(e,e'p)$
scattering depends on the alignment of the deuteron. We study
the resulting FSI effects in the tensor analyzing power in detail
and find substantial FSI effects starting at still low missing
momentum $p_m \gsim 0.9 fm^{-1}$. At larger $p_m \gsim 1.5 fm^{-1}$,
FSI completely dominates both missing momentum distribution
and tensor analyzing power. We find that to a large extent FSI
masks the sensitivity of the tensor analyzing power to
models of the deuteron wave function. For the transversely
polarized deuterons the FSI induced forward-backward asymmetry of
the missing momentum distribution is shown to have a node at
precisely the same value of $p_m$ as the PWIA missing momentum
distribution. The position of this
node is not affected by FSI and can be a tool to
distinguish experimentally between different models for the
deuteron wave function.


\newpage
\doublespace

\section {Introduction}

The deuteron being the simplest nucleus, the experimental
study of the short distance (i.e. large momentum) behaviour
of the deuteron wave function is of particular importance
for understanding the short range properties of the nuclear
forces. The $^2H(e,e'p)$ scattering at large $Q^2$ and large
missing momentum is one of
the reactions with potential sensitivity to the short-range
structure of the deuteron. Indeed, in the PWIA (plane wave
impulse approximation) the $^2H(e,e'p)$ cross section is well
known to probe the deuteron ground state
momentum distribution $\vert \Psi(\vec k)\vert^2$,
since in PWIA the missing momentum $\vec p_m$ can be
identified with the initial $pn$ relative momentum $\vec k$
(for a review see \cite{frulmo}).
However, this simple relation between the deuteron wave
function and the missing momentum distribution is affected
by final state interaction (FSI) between the struck proton and
the spectator neutron.
Despite the deuteron being a very dilute target,
in our recent study \cite{deu} of unpolarized
$^2H(e,e'p)$ scattering at large
$Q^2$, we showed that
FSI effects overwhelmed the PWIA distribution at
$p_m$ $\gsim$ 1.5 fm$^{-1}$. (A similar effect was found
for $^4He$ \cite{hel} and heavier nuclei \cite{4,5}.)
There is a basic necessity to disentangle the short-distance
and/or large relative momentum wave function of the deuteron
and to this end, one can think of exploiting the tensor
polarization of the deuteron. Indeed,
due to the mixing of the S- and D-wave components by the tensor
nuclear forces, the probability of FSI between the
struck proton and the spectator neutron depends on the alignment
(tensor polarization) of the deuteron. (Hereafter the
polarization axis is chosen along the $(e,e')$ momentum
transfer $\vec q$, while
$\theta$ is the angle between $\vec q$ and $\vec p_m$.)
Similar sensitivity to tensor polarization of the double scattering
amplitude in elastic $p$ - $^2 \vec H$ scattering was discussed earlier
\cite{harri,fragla}.
The $^2H(e,e'p)$ experiments with both polarized and unpolarized
deuterons constitute an important part of the program at CEBAF
\cite{cebaf}. The tensor analyzing power can also be measured in the
$^2 \vec H(e,e'p)$ reaction on the
polarized deuteron jet target in the HERMES experiment at HERA
\cite{hera}.

If only the deuteron is polarized,
a key measurable quantity is the tensor analyzing power
\arr
A\ \equiv\ {{\sigma_++\sigma_--2\sigma_0}
\over{\sigma_+ + \sigma_- + \sigma_0}}
\ =\ 2{{\sigma_+-\sigma_0}\over{2\sigma_++\sigma_0}} \, ,
\label{eq:1}
\endarr
where $\sigma_{\mu}$ is the cross section for
the polarization state of the deuteron
$\mu = \pm 1, 0$
and we use the property $\sigma_+ = \sigma_-$.
By definition, $ -2 \leq A \leq 1$, the limits being saturated
when $\sigma_+ = 0$ and $\sigma_o = 0$, respectively.
Already in PWIA the tensor analyzing power $A$ is
nonzero, as the D-wave admixture in the deuteron produces
the tensor polarization dependent deviations from the
spherical shape of the deuteron wave function.
The interest in $A$ stems from the potential
sensitivity to the not so well known D-wave component of the deuteron
and from the possibility of discrimination among
different models for the deuteron ground state.
For the same reason of the deviation from a spherically
symmetric shape, the probability of FSI also depends on the
tensor polarization: at large $\vec q$ the struck proton
flies in the direction of $\vec q$, the probability of
interaction with the spectator neutron is higher if the
deuteron is aligned along $\vec q$ and the alignment
depends on the tensor polarization.
Consequently, the interpretation of the experimental data
on the tensor analyzing power requires a quantitative
understanding of FSI effects.

In this paper we present a detailed Glauber theory description of
FSI effects in the  $^2 \vec H(e,e'p)$ scattering on tensor polarized
deuterons. In section 2 we discuss PWIA predictions for the
tensor analyzing power. Interpretation of the rich structure of the
tensor analyzing power in terms of nodal properties and interference
of the S- and D-wave amplitudes is presented in section 3.
We point out a potential sensitivity of the tensor analyzing power
to models of the deuteron wave function. In section 4 we introduce
the Glauber theory description of FSI and describe FSI modifications
of S- and D-wave amplitudes, in section 5 we present our main results
for FSI effects. We find strong FSI, which leads to complex
modifications of the tensor analyzing power and of the missing momentum
distribution for different polarizations of the deuteron.
FSI effects are particularly strong at large $p_m$ and are an
important background to the extraction of properties of the
short-distance $np$ interaction.
Still another FSI effect - the forward-backward asymmetry of the
missing momentum distribution - is discussed in section 6.
Remarkably, this forward-backward asymmetry probes a nodal structure
of certain undistorted S- and D-wave amplitudes which is free of
uncertainties from FSI. In the conclusions section we summarize
our main results. Numerically, FSI effects in the tensor
analyzing power turn out to be substantial already at
$p_m$ $\gsim$ (0.9-1.0)fm$^{-1}$ and to a large extent mask the
potential sensitivity to models of the deuteron wave function.

\section {Tensor analyzing power in plane wave impulse approximation}

In this communication we would like to concentrate on FSI effects
and to focus on the simpler case of the longitudinal response.
For this case, the photon can be treated as a scalar operator
which does not change the spin state of the proton and the neutron
(for the full hadronic tensor in deuteron scattering, see
\cite{arhoe}). Then, factorizing out the electromagnetic current
matrix element of the struck proton, the reduced nuclear
amplitude for the exclusive process $^2H(e,e'p)n$ is given by
\cite{frulmo,deforest,boffi}
\arr
{\cal M}_{\nu \mu}=\int d^{3}\vec{r}
\exp(i\vec{p}_{m}\vec{r}\,)\chi_{\nu}^{*}S_{\nu}(\vec{r}\,)
\Psi_{\mu}(\vec{r}\,),
\label{eq:2}
\endarr
where
$\vec r$ $\equiv$ $\vec r_n - \vec r_p$,~
$\Psi_{\mu}(\vec{r})$ is the
wave function
for the deuteron spin state $\mu$ and $\chi_{\nu}$ stands for the spin wave
function of the final $pn$ state. The struck proton is detected
with momentum $\vec{P}$,~ $\vec{q}$ is the $(e,e')$ momentum
transfer and
$\vec{p}_m$ $\equiv$ $\vec P - \vec q$ is the missing
momentum. $S_{\nu}(\vec{r})$ is the $S$-matrix of FSI between
the struck proton and the spectator neutron,
which describes the distortion of the outgoing waves.
The subscript $\nu$ is a reminder that the $np$ interaction in the
final state may depend on the spin state.
The observed momentum distribution corresponding to
a given deuteron polarization $\mu$ is the sum
$W_{\mu}(\vec p_m) = \displaystyle {\frac{1}{(2 \pi)^3}}
\sum_{\nu} \vert {\cal{M}}_{\mu \nu} \vert ^{2}$.
The deuteron wave function for states $\mu = \pm,0$ is \cite{bonn}
\arr
\Psi_+  = \Big(\frac{u(r)}{r}Y_{00}(\hat r)+
\sqrt{\frac{1}{10}} \frac{w(r)}{r}Y_{20}(\hat r)
\Big)\vert 1\rangle - \sqrt{\frac{3}{10}}
\frac{w(r)}{r} Y_{21}(\hat r)
\vert 0\rangle + \sqrt{\frac{3}{5}}\frac{w(r)}{r} Y_{22}(\hat r)
\vert  -1\rangle
\label{wf+}
\endarr

\arr
\Psi_0  =
\Big(\frac{u(r)}{r}Y_{00}(\hat r)-
\sqrt{\frac{2}{5}} \frac{w(r)}{r}Y_{20}(\hat r)
\Big)\vert 0\rangle
+\sqrt{\frac{3}{10}}
\frac{w(r)}{r} Y_{2-1}(\hat r) \vert 1\rangle
+ \sqrt{\frac{3}{10}}\frac{w(r)}{r} Y_{21}(\hat r)
\vert -1\rangle
\label{wf0}
\endarr
where $u/r$ and $w/r$ are the $S$ and $D$-wave radial wave functions
of the deuteron, with the normalization $\int dr (u^2+w^2) = 1$.
Here, $\vert \sigma \rangle $ stands for the initial
triplet spin wave function with the projection $\sigma = 0, \pm 1$.

It is convenient to introduce the S- and D-wave amplitudes
\bea
s(\vec p_m) &\equiv&
\int d^3 \vec r \, \frac{u(r)}{r} \, Y_{00}(\hat r) \, S(\vec r) \,
\exp(i \vec p_m \vec r) \label{ms} \\
d_{\tau} (\vec p_m) &\equiv&
\int d^3 \vec r \, \frac{w(r)}{r} \, Y_{2 \tau}(\hat r) \,  S(\vec r) \,
\exp(i \vec p_m \vec r) \,, \label{md}
\eea
where $\tau = 0, \pm 1, \pm 2$ and we suppressed the possible weak
dependence of the FSI factor $S(\vec r)$ on the relevant spin
state $\vert \sigma \rangle$, see below. In terms of these
amplitudes the momentum distributions can be written as
\bea
W_+(\vec p_m) & = & \frac {1}{(2 \pi)^3} \left (
\vert s (\vec p_m) + \sqrt{ \frac{1}{10}} d_o (\vec p_m) \vert ^2
+  \frac{3}{10} \vert d_1 (\vec p_m) \vert ^2
+  \frac{3}{5} \vert d_2 (\vec p_m) \vert ^2 \right )
\label{wpexpl} \\
W_o(\vec p_m) & = & \frac {1}{(2 \pi)^3} \left (
\vert s (\vec p_m) - \sqrt{ \frac{2}{5}} d_o (\vec p_m) \vert ^2
+ \frac{3}{10} \vert d_{-1} (\vec p_m) \vert ^2
+ \frac{3}{10} \vert d_1 (\vec p_m) \vert ^2 \right )
\label{w0expl}   \, .
\eea
In order to set up the framework, we first cite simple results
for PWIA, where $S(\vec r) = 1$.
Denoting
\arr
s(p_m)\ \equiv\ \int dr r  u(r) j_o(p_mr),\ \
d(p_m)\ \equiv\ \int dr r w(r) j_2(p_mr) \, ,
\label{eq:8}
\endarr
we find in PWIA
\arr
s^{PWIA}(\vec p_m) = \sqrt{4 \pi} s(p_m) \label{spwia} \\
d^{PWIA}_{\tau}(\vec p_m) = - 4 \pi Y_{2 \tau} (\hat p_m) d(p_m)
\label{dpwia}  \,.
\endarr
Denoting $W^{PWIA}_{\mu} (\vec p_m)
= N_{\mu}(\vec p_m)$, we have
\arr
N_{\pm}(p_m,\theta)  = N (1 + \frac{1}{2} A)
\label{eq:6}
\endarr

\arr
N_{o}(p_m,\theta)
& \hspace{-3mm}=& \hspace{-3mm} N (1 - A)
\label{eq:7}
\endarr
where
\arr
N = \frac{1}{3} (N_+ + N_- + N_o)
= \frac{1}{2 \pi^2} \left( s^2(p_m) + d^2(p_m) \right)
\endarr
is the missing momentum distribution for the unpolarized deuteron and
the tensor analyzing power equals
\arr
A^{PWIA}(p_m,\theta)\ =\ \frac{1}{2}
\frac {\sqrt{8} s(p_m)d(p_m)+ d^2(p_m)}
{d^2(p_m)+s^2(p_m)}
\bigg(1-3cos^2(\theta)
\bigg) \,.
\label{eq:9}
\endarr
Notice the $\propto (1-3 \cos^2(\theta))$ behaviour of
$A^{PWIA}(\vec p_m)$, by which
\arr
A^{PWIA} (p_m, \cos(\theta) = \pm \frac{1}{\sqrt{3}}) = 0
\label{pwpro1}
\endarr
and
\arr
A^{PWIA}(p_m, \theta =0^o, 180^o) = -2 A^{PWIA}(p_m, \theta=90^o) \,.
\label{pwpro2}
\endarr
Also, notice the property
\arr
N_{0} (p_m, \theta = 90^o) =
N_{\pm}(p_m, \theta = 0^o,180^o)
 = \frac{1}{2 \pi ^2} \left ( s(p_m) - \frac{1}{\sqrt{2}} d (p_m)
\right ) ^2
 \,,
\label{pwpro3}
\endarr
which holds because $Y_{2 \pm 1} (\hat p_m) = 0$ and therefore
$d_{\pm 1} (\vec p_m) = 0$ for $\theta = 0^o,
90^o, 180^o$ and $Y_{2 \pm 2} (\hat p_m) = 0$ and therefore
$d_{\pm 2} (\vec p_m) = 0$ for $\theta = 0^o, 180^o$.

\section{Nodes of S- and D-wave amplitudes, S-D wave interference
and diffractive dip-bump structure in PWIA}

The salient features of the PWIA distributions $N_{\mu}(\vec p_m)$ and
the tensor analyzing power $A^{PWIA}(\vec p_m)$ can easily be
understood in terms of the $p_m$-dependence of $s(p_m)$ and $d(p_m)$.
For the sake of definiteness, consider transverse kinematics, i.e.
$\theta = \frac{\pi}{2}$.
The separate S- and D-wave function contributions (dotted line and
dash-dotted line, respectively) to $N_o(p_m, \theta = 90^o)$
(solid line) and $N_+(p_m, \theta = 90^o)$ (solid line)
are shown in Figs.1c,d, respectively,
for the Bonn potential \cite{bonn}.
Notice a stronger D-wave contribution to $N_+( p_m, \theta = 90^o)$.
The total distributions $N_{\mu} (p_m)$
include also the S-D interference terms not shown separately.
The S-wave amplitude $s(\vec p_m)$ has a zero at
$p_m = p_m^{(1)} = 2.22 fm^{-1}$ for the Bonn potential.
The D-wave amplitudes
$d_{\tau} (\vec p_m)$ have a kinematical zero at $p_m = 0$
and are node-free in the considered
region of $p_m$. The S- and D-wave amplitudes become comparable
at $p_m \gsim 1.5 fm^{-1}$, and the D-wave contribution takes over
at larger $p_m$. For the Bonn potential $N_o(p_m, \theta = 90^o)$
has a zero at $p_m = p_m^{(2)}=1.58 fm^{-1}$ for the destructive
interference and exact cancelation of the S-wave
and the D-wave contributions, when $d(p_m) = \sqrt{2} s(p_m)$, see
Eq. (\ref{pwpro3}). At $p_m > p_m^{(1)}$, the S-D interference
in $N_o( p_m, \theta = 90^o)$ becomes constructive. The
$p_m$-dependence of $N_o(p_m, \theta = 90^o)$ and
$N_+(p_m, \theta = 90^o)$ is strikingly different. For the
constructive S-D interference, $N_+(p_m, \theta = 90^o)$ shown in
Fig. 1d  exhibits a smooth, structureless decrease with $p_m$.
The effect of the dip in the S-wave contribution is completely
masked by an overwhelming D-wave contribution. Notice the destructive
S-D interference in $N_+( p_m, \theta = 90^o)$ at $ p_m > p_m^{(1)}$.

These properties of $s(p_m)$ and $d(p_m)$ have simple manifestations
in the tensor analyzing power.
At $p_m \to 0$ we have
$A^{PWIA}(p_m, \theta=90^o) \to 0$ for the vanishing of the
D-wave function $d(p_m)$. With increasing $p_m$,
$A^{PWIA} (p_m, \theta=90^o)$ (dotted line in Fig.2a)
stays positive, and saturates
at $p_m = p_m^{(2)}$ the upper bound $A^{PWIA}(p_m,
\theta=90^o) = 1$, when $N_o(p_m,\theta=90^o)$ vanishes.
With the further increase of $p_m$, the tensor analyzing power
decreases and takes the value $A^{PWIA}(p_m,\theta=90^o)
= \frac{1}{2}$ when the S-wave function $s(p_m)$ has a node
at $p_m = p_m^{(1)} = 2.22 fm^{-1}$ for the Bonn potential.
Beyond the node, the S-wave function rises in magnitude,
leading to a change of sign of $A^{PWIA}(p_m, \theta=90^o)$
at $p_m = p_m^{(3)} = 2.73 fm^{-1}$ for the Bonn potential,
when $s(p_m) = - \frac{1}{\sqrt{8}} d(p_m)$.

The position $p_m^{(2)}$ of the zero in
$N_{0} (p_m, \theta = 90^o)$ and $N_{\pm}(p_m, \theta = 0^o,180^o)$,
and the position $p_m^{(3)}$ of the zero of
$A^{PWIA}(p_m, \theta=90^o)$, are a matter of delicate cancelations
between the S- and D-wave functions. As such, they are sensitive
to the D-wave function, which remains a not so well known quantity
in the theory of the deuteron. In Fig. 2a we compare the predictions
for $A^{PWIA}(p_m, \theta=90^o)$ from the Bonn and Paris \cite{paris}
wave functions (dotted and dash-dotted curve, respectively).
The qualitative behaviour of the results is the same,
but for the Paris potential, $N_{0} (p_m, \theta = 90^o)$ has a
zero at a value of $p_m^{(2)} =  1.50 fm^{-1}$, which is
$0.08 fm^{-1}$ smaller than for the Bonn wave function.
The sensitivity to the model of the wave function increases with
$p_m$: for the Paris wave function, the tensor analyzing power
changes sign at a value of $p_m^{(3)} = 2.52 fm^{-1}$,
by about $0.2 fm^{-1}$ smaller than for the Bonn wave function.

The angular dependence of $N_{\mu}(p_m, \theta)$
in PWIA for different missing momenta
is shown in Fig.3 and the angular distribution of $A^{PWIA}(p_m,
\theta)$ is shown by the dotted curve in Fig.4.
By virtue of (\ref{eq:6}) and (\ref{eq:7})
a minimum in $N_+$ (dotted curve)
corresponds to a maximum in $N_o$ (dashed curve) and vice versa,
and $N_+ = N_o$ at $\theta = \arccos \left ( \pm \frac{1}{\sqrt{3}}
 \right ) = 90^o \pm 35.3^o$.
By virtue of the definition (\ref{eq:1}) the sign of the difference
between $N_{\pm}(p_m, \theta)$ and $N_o(p_m, \theta)$
depends on the tensor analyzing power $A$. For instance,
$A^{PWIA}(p_m, \theta = 90^o)$ changes sign at
$p_m^{(3)} = 2.73 fm^{-1}$ (see Figs.2,4) and for
$p_m > p_m^{(3)}$ we have $ N_+(p_m, \theta = 90^o) <
N_o(p_m, \theta = 90^o)$, whereas for
$p_m < p_m^{(3)}$ we have $N_+(p_m, \theta = 90^o) >
N_o(p_m, \theta = 90^o)$, cf. Figs. 3a-e and Fig. 3f.
In PWIA, $N_{\mu} (p_m, \theta)$ are even functions of $\cos \theta$,
and there is no forward-backward asymmetry: $N_{\mu} (p_m, \theta = 0^o)$
 = $N_{\mu} (p_m, \theta = 180^o)$.

\section{FSI effects and nodal structure of S- and D-wave amplitudes}

FSI effects substantially modify the above pattern.
In this work we focus on large $Q^{2}\gsim$ (1-2)\,GeV$^{2}$
 of interest in the
CEBAF experiments, where the kinetic energy of the struck proton
$T_{kin}\approx Q^{2}/2m_{p}$ is high and FSI can be described by
Glauber theory \cite{glau}. The onset of color transparency
effects at the highest $Q^2$ attainable at CEBAF will be discussed
elsewhere. Defining transverse and longitudinal
components $\vec{r}$ $\equiv$ $(\vec{b}+z\hat q)$,
where $\hat q $ denotes the unit vector in $z$-direction
and $\vec b$ is the impact parameter, we can write
\beq
S_{\nu}(\vec{r}) = 1-\theta(z)\Gamma_{\nu}(\vec{b}),
\label{eq:3}
\endeq
where $\Gamma_{\nu}(\vec{b})$ is the profile function of the
proton-neutron scattering, which at high energy can conveniently
be parameterized as
\beq
\Gamma_{\nu}(\vec{b}) \ \equiv\
{ \sigma_{tot}^{(\nu)} (1 - i \rho_{\nu}) \over 4 \pi b_{o}^2  }
\exp \Big[-{\vec{b}^2 \over 2 b_{o}^2} \Big]
 = (1 - i \rho_{\nu}) G_{\nu}(\vec b)
\label{eq:4}
\endeq
($\rho$ is the ratio of the real to imaginary part of the
forward $np$ elastic scattering amplitude, and $b_o^2$ is the
diffraction slope). The step-function $\theta(z)$
in eq.(\ref{eq:3}) tells that
the FSI vanishes unless the spectator neutron is in the
forward hemisphere with respect to the struck proton.
A thorough test of the Glauber theory in the scattering of
2 GeV polarized deuterons on protons has been performed in \cite{13},
higher energy data on elastic $pd$ scattering are discussed in
\cite{14}, a generic review on the Glauber theory analysis of
nucleon-nucleus scattering is given in \cite{15}.
For the high energy and momentum of struck protons and
for the simple case of the longitudinal response considered here,
the spin state of the proton and neutron does not change and
FSI between struck proton and spectator neutron proceeds
in the spin triplet state. What enters the Glauber FSI factor
(\ref{eq:4}) is the $np$ total cross section $\sigma_{tot}^{(\nu)}$
for the triplet state with polarization $\nu$.
We can express $\sigma_{tot}^{(\nu)}$
in terms of the conventional longitudinal, $ \Delta \sigma_L$,
and transverse, $\Delta \sigma_T$, cross section differences
\cite{biry,lll}
\arr
\sigma_{tot}^{(\nu)} = \sigma_o - \frac{1}{2} \Delta \sigma_L
(p_+ +  p_- - p_o) - \Delta \sigma_T p_o \, ,
\label{cspin}
\endarr
where $p_{\nu}$ indicates the probability of
the spin state $ \vert \nu \rangle$, $p_+ + p_- + p_o = 1$.
The $pn$ cross sections for the pure state $\nu$ are obtained
by putting either $p_+ + p_- =1$ or $p_o = 1$
in eq. (\ref{cspin}).
The experimental data on $\Delta \sigma_L$ and $\Delta \sigma_T$,
reviewed in \cite{lll}, show that $\Delta \sigma_L$ and $\Delta
\sigma_T$ are $\lsim (10-20) \%$ of the unpolarized cross section
$\sigma_o$. They exhibit an interesting energy dependence in the
region $0.5 GeV \lsim T_{kin} \lsim 1 GeV$ and tend to decrease
at higher $T_{kin}$.
In principle, the experimental isolation of FSI effects and
investigation of their $Q^2$ ($T_{kin}$) dependence can shed some
light on the energy dependence of $\sigma_{tot}^{(\nu)}$.
Unfortunately, the direct experimental measurements of $\rho_{\nu}$
and of the diffractive slope $b_o^2$ for pure spin states $\nu$
are not yet available. The interplay of FSI and tensor polarization
effects leads to a very complex pattern of the $(p_m, \theta)$
dependence of momentum distributions. In order to emphasize the
salient features of FSI effects, here we omit small corrections
due to $\Delta \sigma_L, \Delta \sigma_T$ and, for the sake of
definiteness, consider $\sigma_{tot} \approx 40mb$, $\rho\approx
-0.4$ and $b_{o}\approx 0.5fm$, appropriate for unpolarized
$np$ scattering at $T_{kin} \sim 1 GeV$ ($Q^2 \sim 2 GeV^2$)
\cite{15,lll,16}. The possibility of using FSI effects to pin
down the energy dependence of $\Delta \sigma_L$ and $ \Delta
\sigma_T$ will be discussed elsewhere.

Let us decompose the missing momentum $\vec{p}_{m}$ into the transverse and
longitudinal components $\vec{p}_{m}=\vec{p}_{\perp}+p_{z}\hat{q}$.
The radius $b_{0}$ of FSI is
much smaller than the radius of the deuteron $R_{D} \sim 2$fm.
Consequently, as compared to $u(r)$ and $w(r)$, the FSI operator
$\theta(z)\Gamma(\vec{b})$ is a "short-ranged" function of
$\vec{b}$ and a "long-ranged" function of $z$, and
this anisotropy of $S(\vec r\,)S^{\dagger}(\vec r\,')$ leads
to the angular anisotropy of $W(\vec{p}_{m})$ \cite{deu,hel}.
Another striking effect
is the forward-backward asymmetry
$W(p_{\perp},-p_{z})\neq W(p_{\perp},p_{z})$, which
has its origin in
the nonvanishing real part of the $p$-$n$ scattering amplitude
$\rho\neq 0$.
This angular anisotropy of FSI effects leads to quite a complex
interplay of FSI and S-D interference,
breaking of the PWIA symmetry properties (\ref{pwpro1})
- (\ref{pwpro3}) and strong departure from the PWIA law
$A \propto (1 - 3 \cos ^2 (\theta))$. A further complication
comes from the finite $\rho$, which, apart from the forward-backward
asymmetry leads to complex-valued S- and D-wave amplitudes
and to the filling of the dips in $N_{\mu}(\vec p_m)$.

The gross features of FSI effects can be understood, though, in terms
of FSI distortions of the S- and D-wave amplitudes (\ref{ms},\ref{md}).
The main findings of our previous study of the unpolarized
$^2H(e,e'p)$ scattering can be summarized as follows:
\begin{enumerate}
\item Angular anisotropy of the FSI factor (\ref{eq:3})
leads to an angular dependence of the S- and D-wave amplitudes
(\ref{ms},\ref{md}) which is different from the PWIA expressions
(\ref{spwia},\ref{dpwia}).
\item At small $p_m \lsim 0.5 fm^{-1}$, FSI reduces $s(\vec p_m)$
by $\sim 3.5 \%$. In this region of small $p_m$, the FSI
reduction is to a good approximation $\theta$- independent.
It describes the attenuation of the flux of struck protons
due to inelastic $np$ interactions and the flow of events from
small to large $p_{\perp}$ due to elastic $np$ rescattering.
\item Destructive interference between PWIA and FSI
components of $s(\vec p_m)$ produces a node of the real part
of $s(\vec p_m)$, shown by the solid curve in Fig.5,
in transverse kinematics, $\theta = 90^o$,
at $p_{\perp} = p_m^{(4)} = 1.37 fm^{-1}$ (for the Bonn wave
function). The presence of this node, shown in Fig.5, is solely
a consequence of the PWIA-FSI interference, it is not related
to the presence of the node in the PWIA amplitude (\ref{spwia})
at much larger $p_m = p_m^{(1)} = 2.22 fm^{-1}$.
The position of the node of $Re s(\vec p_m)$
in $p_{\perp}$ only weakly depends on the longitudinal
component $p_{z}$ of the missing momentum.
\item At larger $p_{\perp}$, the FSI contribution to $s(\vec p_m)$
takes over the PWIA amplitude, and leads to a sharp
peak at $\theta = 90^o$. This peak has its origin in elastic
rescattering of the struck proton on the spectator neutron,
which contributes predominantly to $\theta = 90^o$ for the reason
that in high-energy elastic rescattering the momentum transfer
is predominantly a transverse one. Fig.3 shows how this $90^o$ peak
builds up with the increase of $p_m$.
\end {enumerate}

Distortion of the D-wave amplitudes $d_{\tau}(\vec p_m)$
has a certain specifics due to the fact that the radius of
$pn$ FSI is much smaller than the radius of the deuteron.
For crude estimates, consider the approximation
\arr
\Gamma (\vec b) \approx \frac{1}{2} \sigma_{tot}(pn)
(1 - i \rho) \delta(\vec b) \,.
\label{gamaprox}
\endarr
For the presence of $\delta (\vec b)$ in
(\ref{gamaprox}), FSI contributions to
$d_{\tau}(\vec p_m)$
will be dominated by $\cos \theta = 1$, where $Y_{2 \pm 2}(\theta)
=Y_{2 \pm 1} (\theta) = 0$. This leads to an extra, and
very strong kinematical suppression of FSI contributions to
$d_{ \pm 2}(\vec p_m)$ and $d_{\pm 1} (\vec p_m)$.
On the other hand, the FSI contribution to $d_o (\vec p_m)$ does not
have any special suppression and, in close similarity to
an S-wave amplitude, the real part of
$d_o(\vec p_m)$, shown by the dotted curve in Fig.5,
develops a node at $p_{\perp} = p_m^{(5)}
= 2.87 fm^{-1}$ (for the Bonn wave function).
Compared to $d_o (\vec p_m)$, in $d_{\pm 1} (\vec p_m)$ and
$d_{\pm 2} (\vec p_m)$ FSI effects are negligible in the region
of $p_m \lsim (3-4) fm^{-1}$ of practical interest.
For instance, $Re \, d_2 (\vec p_m)$, shown by the dashed curve
in Fig.5 does not develop a node.

Still another striking effect of FSI is a nonvanishing
$d_{\tau}(\vec p_m)$
at $\vec p_m = 0$, as shown in Fig.5 for $\theta = 90^o$.
Indeed, in the momentum space, the FSI contribution can be
represented as a convolution of the PWIA amplitude with a
certain smearing function. Because of this smearing,
$d_{\tau}^{FSI}(\vec p_m = 0) \not= 0 $.
Because of the kinematical suppression of the PWIA amplitude at
small $p_m$, even a weak FSI amplitude which has a sign opposite
to that of the PWIA amplitude, leads to a zero of
$Re \, d_{\tau}(\vec p_m)$ at small $p_m$. Fig. 5 shows how
$Re \, d_o(p_m, \theta = 90^o)$ develops such a zero at $p_{\perp}
= p_m^{(6)} = 0.09 fm^{-1}$ for $\theta = 90^o$.
The nonvanishing D-wave contribution at $\vec p_m =0$ when FSI
is included is evident also in Figs.1a,b. However, because of
the strong inequality $\vert d_{\tau}(\vec p_m) \vert \ll
\vert s(\vec p_m)\vert $ in the vicinity of $p_m = 0$,
it is very difficult
to observe this fine structure of the D-wave amplitudes
experimentally.

\section{FSI effect on S-D interference and position of
diffractive dips}

Now we turn to the interpretation of our results in view
of the above described properties of FSI.
At small $p_m \lsim 0.5 fm^{-1}$, both $W_o(\vec p_m)$ and
$W_+(\vec p_m)$ are dominated by the S-wave contribution and are
reduced by $\simeq 7 \%$ with respect to the PWIA distributions
$N_o (\vec p_m)$ and $N_+ (\vec p_m)$.
Let us again start the discussion of larger $p_m$ with the simpler
case of $W_o (\vec p_m, \theta = 90^o)$ shown by the solid curve
in Fig.1a. The S-wave contribution,
shown by the dotted line, has a dip at $p_m = p_m^{(4)} =1.37 fm^{-1}$
for the node of $Re \, s(\vec p_m)$ (the same dip shows up in
$W_+ (\vec p_m)$ shown by the solid line in Fig.1b).
In the PWIA case, the dip of the S-wave contribution
to $N_{\mu} (\vec p_m)$  was located
at $p_m = p_m^{(1)} =2.22 fm^{-1}$ and its effect on $N_{\mu}
(\vec p_m)$ was barely observable for the overwhelming D-wave
contribution. The dip in the S-wave component of $W_{o,+}
(\vec p_m)$ is filled in by the contribution from the imaginary part
of the S-wave amplitude.
The D-wave component of $W_o (\vec p_m)$ (dash-dotted curve
in Fig.1a) shows an irregularity
at $p_m = p_m^{(5)} =2.87 fm^{-1}$ due to the node of $Re \, d_o (\vec
p_m)$. The possible dip of the D-wave component of $W_o (\vec
p_m)$ is filled in by a contribution from $Im \, d_o (\vec p_m)$
(and also by the much smaller FSI components of $d_2 (\vec
p_m), d_{\pm 1} (p_m)$).
This irregularity is difficult to observe experimentally for
the predominant S-wave contribution to $W_o(p_m, \theta = 90^o)$.

Because of the shift of the zero of $Re \, s(\vec p_m)$ to
$p_m = p_m^{(4)} = 1.37 fm^{-1}$ compared to
$p_m^{(1)} = 2.22 fm^{-1}$ in the PWIA,
the condition $Re \, s (p_m, \theta = 90^o) \simeq \frac{1}{\sqrt{2}}
Re \, d(p_m, \theta = 90^o)$ and the resulting dip in $W_o (\vec p_m)$
will be met at $p_m = p_m^{(7)} = 1.21 fm^{-1}$, a strong shift from
$p_m^{(2)} \simeq 1.58 fm^{-1}$ in the PWIA, cf. solid curves in
Figs.1a,c. The dip in $W_o (\vec p_m)$ is partly filled in by
the contribution $ \propto \vert Im \, s (\vec p_m) \vert ^2$
(there is also a small effect of interference between
$Im \, s (\vec p_m)$ and $Im \, d_o(\vec p_m)$ and a small
contribution from nonvanishing FSI components of $d_{\pm 1}$
and $d_{\pm 2}$). FSI effects in
$W_o (p_m, \theta = 90^o)$ are very strong and lead to a drastic
difference between $W_o (p_m, \theta = 90^o)$ and the PWIA
distribution $N_o (p_m, \theta = 90^o)$  starting from a still
small missing momentum $p_m \gsim (0.8-0.9) fm^{-1}$. In contrast to
$W_o (p_m, \theta = 90^o)$, FSI effects in
$W_+ (p_m, \theta = 90^o)$
are much smaller for the reason that the S-D interference is much
weaker and constructive. Still, as compared to a structureless
and monotonously decreasing PWIA
distribution $N_+ (p_m, \theta = 90^o)$, the dip in the S-wave
contribution leads to a substantial structure in
$W_+ (p_m, \theta = 90^o)$ shown by the solid curve in Fig1b.

At a sufficiently large $p_{\perp}$, the contribution of
elastic rescattering to $W_{\mu} (p_m, \theta = 90^o)$ admits
to a certain extent the probabilistic interpretation.
At $p_{\perp} \gsim 2.5 fm^{-1}$ the $90^o$ peak in
$W_+ (p_m, \theta = 90^o)$ is higher than in
$W_o (p_m, \theta = 90^o)$, in agreement with the fact that the
alignment of the proton and neutron along the z-axis is stronger
for $\mu = \pm 1$ than for $\mu = 0$. Indeed, precisely such a
sign of the quadrupole deformation follows from the sign of the
quadrupole moment of the deuteron \cite{harri, fragla}.

The results for the parallel kinematics, $p_{\perp} = 0$,
$\theta = 0^o$ and $\theta = 180^o$, are shown in Fig.6.
At small $p_{\perp}$, FSI amplitudes have been shown \cite{deu}
to be suppressed by a small factor $\propto \displaystyle{
\frac{\sigma_{tot}^{(pn)}}{4 \pi R_D^2} \sim \left ( \frac
{b_o}{R_D} \right ) ^2}$. For this reason of small FSI effects,
$W_+(p_m, \theta = 180^o)$ (dashed curve in Fig.6a)
has a shallow dip and $W_+(p_m, \theta = 0^o)$ (solid curve in Fig.6a)
has a shoulderlike structure at
$p_m$ very close to the position of the dip in the PWIA distribution
$N_+(p_m, \theta = 0^o, 180^o)$ (dotted curve) at $p_m = p_m^{(2)}
= 1.58 fm^{-1}$. Substantial departure from PWIA starts
at $p_m \gsim 1 fm^{-1}$ and is very strong at higher $p_m$.

In Fig.6a, we present also $W_o(p_m, \theta = 90^o)$ (dash-dotted
curve). A comparison with $W_+(p_m, \theta = 0^o)$ (solid curve)
and $W_+(p_m, \theta = 180^o)$ (dashed curve)
shows a dramatic breaking of the PWIA symmetry (\ref{pwpro3})
by FSI effects. Strong breaking of still another PWIA
symmetry, $W_{\mu} (p_m, \theta = 0^o) \not=
W_{\mu} (p_m, \theta = 180^o)$, will be discussed in more
detail below, in section 6.

In spite of the suppression factors $\sim \left ( \frac{b_o}
{R_D} \right )^2$, FSI effects in
$W_+ (p_m, \theta = 0^o,180^o)$ shown in Fig.6a are enhanced for the
destructive S-D interference, see eqs. (\ref{wpexpl},
\ref{w0expl}). In $W_o (p_m, \theta = 0^o,180^o)$
the S-D interference is constructive and consequently,
$W_o (p_m, \theta = 0^o)$ (solid curve) and
$W_o (p_m, \theta = 180^o)$
(dashed curve) are least sensitive to FSI effects.
Even so, FSI effects are noticeable already at relatively small
$p_m \gsim 1 fm^{-1}$, and completely overwhelm
$W_o (p_m, \theta = 0^o,180^o)$ at $p_m \gsim 2 fm^{-1}$.
Notice that at large
$p_m \gsim (2.5-3) fm^{-1}$
the distributions
$W_o (p_m, \theta = 0^o)$ and
$W_o (p_m, \theta = 180^o)$ exhibit a strong excess over, and
decrease very slowly compared to, the PWIA distribution
$N_o (p_m, \theta = 0^o)$ (dotted curve).
The origin of this excess lies in the
$\theta$-function  which emerges in the idealized form of the
Glauber FSI operator (\ref{eq:4}). For the discussion of the
sensitivity of this excess to modifications of the FSI operator
(\ref{eq:4}) for the finite size of the nucleus we refer to
\cite{deu,hel}.

A comment on the interpretation of the above results in terms
of the nuclear transparency ratio $ T_{\mu} = \displaystyle{
\frac{W_{\mu} (p_m, \theta)} {N_{\mu}(p_m, \theta)}}$ is in order.
Only at small $p_m \sim 0$, the result that $T_{\mu}(p_m,
\theta) \simeq 0.93 < 1$ can be interpreted as a conventional
nuclear attenuation. At higher $p_m$, the nuclear transparency
ratio has a strongly polarization dependent dip-bump structure.
The rich and complex $p_m$ and $\theta$
dependence of the nuclear transparency
$T_{\mu} (p_m, \theta)$ is evident from the results shown in
Fig.3. For instance, in transverse kinematics $T_o(p_m^{(7)},
\theta = 90^o) \ll 1$, which is followed by a rapid rise of the
transparency ratio and an infinite peak
$T_o(p_m^{(2)},\theta = 90^o) = \infty$ at the dip of the PWIA
distribution $N_o(p_m, \theta = 90^o)$.
At still larger $p_{\perp}$, in the region dominated by elastic
rescattering effects, we find very strong nuclear enhancement
(antishadowing) $T_{\mu} (p_m, \theta) \gg 1$ (see also
\cite{hel,5}).

The $(p_m, \theta)$ dependence of the tensor analyzing power
naturally follows from the above described properties of the
S- and D-wave amplitudes and $W_{\mu} (p_m, \theta)$.
Fig.4 shows a strong departure of $A(p_m, \theta)$ (solid
curve) from the PWIA
law $\propto (1 - 3 \cos ^2 \theta)$. In the PWIA (dotted curve),
the tensor
analyzing power has two zeroes at $\theta = 90^o \pm 35.3^o$.
The results shown in Fig.4 demonstrate how the FSI distortions
change both the location and number of zeroes of $A(p_m, \theta)$.
(The origin of the forward-backward asymmetry of $A(p_m, \theta)$
will be discussed in more detail below.)
The $p_m$ dependence of $A(p_m, \theta)$ derives from the $p_m$
dependence of the S- and D-wave amplitudes. In contrast to the
PWIA case, $A(p_m, \theta )$ does not vanish at $p_m = 0$, but
numerically  $ \vert A(p_m = 0, \theta) \vert \ll 1$ and is hardly
measurable.

For the sake of definiteness, let us consider in more
detail transverse kinematics $\theta = 90^o$. With increasing
$p_m$, $A(p_m, \theta)$ rises and has a maximum at $p_m \approx
p_m^{(8)} = 1.19 fm^{-1}$,
in the region of the dip in $W_o (p_m, \theta = 90^o)$.
In the opposite to $A^{PWIA} (p_m^{(2)}, \theta = 90^o)$ in
PWIA, $A(p_m^{(8)}, \theta = 90^o)$ does not saturate
the bound $A = 1$. With the further increase of $p_m$, we
encounter  $A(p_m, \theta = 90^o) = \frac {1}{2}$ at $p_m^{(9)}
= 1.42 fm^{-1}$, very close to the position of the FSI induced node
in $Re \, s( \vec p_m)$. At still higher $p_m$, $A(p_m, \theta = 90^o)$
has two zeroes. The first zero, at $p_m = p_m^{(10)} = 1.70 fm ^{-1}$
is the counterpart of the zero at $p_m = p_m^{(3)} = 2.73 fm^{-1}$
in PWIA, caused by the
cancelation of S- and D-wave amplitudes when $ d(p_m) \simeq
- \sqrt{8} s(p_m)$. Such a strong shift of the location of this
zero with respect to PWIA is due to the fact that $s(p_m)$
changes sign much sooner compared to the PWIA case. Compared to the
PWIA case, the exact value of $p_m^{(10)}$ is affected also by different
distortions of $d_{\tau} (\vec p_m)$ and by the contribution from
$Im \, s(\vec p_m), \: Im \, d_{\tau} (\vec p_m)$, see Eqs.
(\ref{wpexpl},\ref{w0expl}).
The origin of the second zero at $p_m = p_m^{(11)} = 2.46 fm^{-1}$
is related to the FSI induced node of $Re \, d_o (\vec p_m)$ at
$p_m = p_m^{(5)} = 2.87 fm^{-1}$.
The inequality $p_m^{(11)} < p_m^{(5)}$ is
an effect of the contribution to $A(p_m, \theta = 90^o)$ from
$\vert d_2 (\vec p_m)\vert ^2$ (the contribution $ \propto \vert
d_1 (\vec p_m) \vert ^2 $ is numerically very small), which makes
$A>0$ at $p_m = p_m^{(5)}$.
At still larger $p_m$, where $W_{\mu} (p_m, \theta=90^o)$ is
dominated by the contribution from elastic $np$ rescattering
in the S-wave, the tensor analyzing power is positive, i.e.
$W_+ > W_o$, for the above discussed larger probability of
rescattering for $\mu = \pm 1$.

Comparison of the PWIA results from the Paris and Bonn wave
functions in Fig.7 shows a significant sensitivity of
$N_{\mu} (p_m, \theta = 90^o)$ to the model of the deuteron wave
function at large momentum. The PWIA tensor analyzing power shown in
Fig.2 exhibits similar model dependence at large $p_m$.
FSI effects mask this sensitivity to the model of the wave function
to a large extent, which naturally derives from the fact that
$\Gamma (\vec b)$ is a short ranged function compared to the
deuteron wave function. For this reason, the FSI component of
S- and D-wave amplitudes becomes fairly insensitive to
details of the large $p_m$ behaviour of the deuteron
wave function.
For instance, whereas in PWIA $N_o(p_m, \theta = 90^o)$ has
a dip at $p_m^{(2)} = 1.50 fm^{-1}$ for the Paris wave function
(dashed curve in Fig.7b)
and at $p_m^{(2)} = 1.58 fm^{-1}$ for the Bonn wave function
(solid curve in Fig.7b), with allowance for FSI a
position of the dip of
$W_o(p_m, \theta = 90^o)$ at $p_m = p_m^{(7)}$ is predicted
very close for the Paris (dashed curve in Fig.7a)
and Bonn (solid curve in Fig.7a) potentials: $p_m^{(7)} = 1.206
fm^{-1}$ and $p_m^{(7)} = 1.208 fm^{-1}$, respectively.
This suggests that the actual value of $p_m^{(7)}$ is much more
sensitive to the strength of the FSI rather than to the wave
function. The difference between $W_+(p_m, \theta = 90^o)$ for the
Paris (dotted curve in Fig.7a) and the Bonn (dash-dotted curve
in Fig.7a) wave functions also becomes marginal compared to the
difference between the PWIA distributions shown in Fig.7b.
Because the position of the dip in $W_o(p_m, \theta = 90^o)$
is mostly controlled by FSI in the spin-triplet state with
$\nu = 0$, one can think of using the high-statistics data on
$^2 \vec H(e,e'p)$ scattering on longitudinal deuterons
in transverse kinematics for the
evaluation of $ \sigma_{tot}^{(o)} = \sigma_o + \frac{1}{2} \Delta
\sigma_L - \Delta \sigma_T$. For instance, if FSI is evaluated
with $\sigma_{tot}^{(o)} = 30 mb$ keeping the same value of $\rho_o$
and $b_o^2$ (at  $T_{kin} \simeq 1 GeV$ the SATURNE data
$\frac{1}{2} \Delta \sigma_L - \Delta \sigma_T \sim 10 mb$
\cite{fon}), then the dip is shifted to $p_m = 1.264 fm^{-1}$
and $1.270 fm^{-1}$ for the Paris and Bonn potentials, respectively.
This shows that there is a residual model dependence of the dip
position and the almost exact coincidence of the values of
$p_m^{(7)}$ evaluated for the Paris and Bonn potentials with
$\sigma_{tot}^{(o)} = 40 mb$ was somewhat accidental.
Still the shift of the dip position from $p_m^{(7)} \simeq
1.21 fm^{-1}$ to $p_m^{(7)} \simeq 1.27 fm^{-1}$ when the $np$ cross
section is changed from $40 mb$ to $30 mb$ is substantially larger
than the difference between the results for Paris and Bonn potentials.
Besides the position of the dip in $W_o(p_m, \theta = 90^o)$,
the large $p_m$ behaviour of $W_{\mu}(p_m, \theta = 90^o)$ is also
sensitive to $\sigma_{tot}^{(o)}$. Namely, according to \cite{deu},
the height of the $\theta = 90^o$ peak in $W_{\mu} (p_m, \theta)$
is proportional to $ \displaystyle { \left ( \frac{\sigma_{tot}^{(o)}}
{4 \pi R_D^2} \right ) ^2}$. However, large $p_m$ behaviour of this
peak is also sensitive to the as yet experimentally not measured
diffraction slope for scattering in this specific spin state.
A more detailed analysis of the sensitivity of the dip position
and of the elastic rescattering peak to
$\Delta \sigma_L$, $\Delta \sigma_T$, and to values of
$\rho$ and $b_o^2$ for pure spin states $\vert \nu \rangle$ goes
beyond the scope of the present paper and will be presented
elsewhere.

\section{Tensor polarization and forward-backward asymmetry}

Above we have seen how FSI masks the effects due to the peculiarities
of the unperturbed wave function of the deuteron.
It is interesting that the forward-backward asymmetry
\arr
A_{FB}^{(\mu)} =
{W_{\mu}(\theta=0^{o},p_{m})-W_{\mu}(\theta=180^{o},p_{m}) \over
 W_{\mu}(\theta=0^{o},p_{m})+W_{\mu}(\theta=180^{o},p_{m})}
\endarr
is a pure PWIA-FSI interference effect caused by the nonvanishing
elastic part of the $p-n$ rescattering amplitude and as such is
proportional to the PWIA amplitudes.
Indeed, compare the amplitudes ${\cal{M}}_{\mu \nu}(\pm p_m)$
for a given initial spin state
$\mu$ at $\theta = 0^o \: (p_z = + p_m)$ and
$\theta = 180^o \: (p_z = - p_m)$.
Making use of $\theta(z)$ = ${1 \over 2}(1+\epsilon(z))$,
 where $\epsilon(z)$ is equal to +1 for $z > 0$ and -1 for $z < 0$,
we have  (we suppress the spin variables $\nu, \mu, \tau $)
\bea
{\cal{M}}(\pm p_m) & = &
\int \Psi(\vec r) \exp(\pm i p_m z)
\bigg(1 - (1-i\rho)G(\vec b)\theta(z)\bigg) d^2bdz  \\
& = &
{\cal{M}}_{PWIA}(p_m) - (1-i\rho)\bigg(F_c(p_m)\pm iF_s(p_m)\bigg)
\nonumber \\
& = &
{\cal{M}}_{PWIA}(p_m) - F_c(p_m) \mp \rho F_s(p_m)
+ i \left ( \rho F_c(p_m)\mp F_s(p_m) \right) \nonumber ,
\eea
where, using the fact that in our case $\Psi (\vec r)$ is an even
function of $z$,
\arr
{\cal{M}}_{PWIA}( \pm p_m) \ \equiv\
\int_{-\infty}^\infty \Psi(\vec r) \exp(\pm i p_m z)
d^2bdz\ =\ 2
\int_0^\infty \Psi(\vec r) \cos(p_m z)
d^2bdz,
\endarr
\arr
F_c(p_m)\ \equiv\
\int_0^\infty \Psi(\vec r)
G(\vec b) \cos(p_m z) d^2bdz,
\label{fcdef}
\endarr
\arr
F_s(p_m)\ \equiv\
\int_0^\infty \Psi(\vec r)
G(\vec b) \sin(p_m z) d^2bdz.
\label {fsdef}
\endarr
Then one obtains
\arr
A_{FB}^{(\mu)} & = & \frac{W_{\mu} (+ p_m) - W_{\mu} (- p_m)}
{W_{\mu} (+ p_m) + W_{\mu} (- p_m)} =
\frac{1}{(2 \pi)^3} \frac{ \sum_{\tau}
\left \{ |{\cal{M}}(p_m)|^2 - |{\cal{M}}( - p_m)|^2 \right \}_{\mu,\tau}}
{W_{\mu} (+ p_m) + W_{\mu} (- p_m)} \nonumber \\
 & = &  - 4 \rho  \frac {1}{(2 \pi)^3} \frac {\sum_{\tau}
 \left \{ {\cal{M}}_{PWIA} (p_m) F_s (p_m)
\right \}_{\mu, \tau}} {W_{\mu} (+ p_m) + W_{\mu} (- p_m)} \, ,
\label{afb}
\endarr
where the sum over $\tau$ is a reminder that for the chosen
deuteron polarization $\mu$ different combinations of $s(\vec p_m)$
and $d_{\tau}( \vec p_m)$ enter the matrix element according to
Eqs. (\ref{wpexpl},\ref{w0expl}).
Because of $F_s (p_m) \to 0$ at $p_m \to 0$, see Eq. (\ref{fsdef}),
$A_{FB}^{(\mu)}$ vanishes at $p_m \to 0$.

In the generic case, both the S-wave and D-wave amplitudes with
$\tau = \pm 2, \pm 1, 0$ contribute to $A_{FB}^{(\mu)}$. The situation
greatly simplifies for transverse deuterons, polarized along the $z$
axis, $\mu = \pm 1$. In this case, D-waves with $\vert \tau \vert = 1,2$
do not contribute to $A_{FB}^{(+)}$, because the corresponding PWIA
amplitudes (\ref{dpwia}) vanish at $\theta = 0^o, 180^o$. Consequently,
only the term with $\tau = 0$ contributes to the numerator in
(\ref{afb}), and the relevant PWIA amplitude equals
\arr
{\cal{M}}_{PWIA}^{(\tau = 0)}
= \sqrt{4 \pi} \left(s(p_m) - \frac{1}{\sqrt{2}} d(p_m)
\right) \, .
\endarr
Therefore, we predict that $A_{FB}^{(+)}$ has a node at precisely
$p_m = p_m^{(2)}$ and the location of this node is not affected by
FSI. This provides a unique possibility of experimental determination
of $p_m^{(2)}$, which is important for testing models of the
deuteron wave function. The sensitivity to the model of the
deuteron is demonstrated in Fig.8b, where we compare $A_{FB}^{(+)}$,
for the Bonn and the Paris wave functions. Notice the very steep
$p_m$-dependence of $A_{FB}^{(+)}$ around the node.
Eqs. (\ref{fcdef},\ref{fsdef}) show that while $F_c(p_m)$ decreases
with $p_m$, $F_s(p_m)$ first rises with $p_m$. As a matter of fact,
in  spite of the FSI suppression factor $\sim (\frac{b_o}{R_D})^2$,
$F_s(p_m)$ is quite large in the vicinity of the node and gives rise
to a peak in $A_{FB}^{(+)}$ at $p_m \approx 1.3 fm^{-1}$.

For the longitudinal deuterons, $\mu = 0$, the corresponding PWIA
am\-pli\-tudes do not have any interesting nodal structure, see Fig.1d,
which shows quite a structure\-less
$N_o(p_m,\theta=0^o,180^o)$ (solid line).
Correspondingly, $A_{FB}^{(0)}$
does not exhibit any structure at $p_m \lsim (3-4) fm ^{-1}$.

\section {Summary and conclusions}

The purpose of the present study has been an analysis of effects
of the quadrupole deformation of the deuteron in $^2 \vec H(e,e'p)n$
scattering on tensor-polarized deuterons. Because FSI is sensitive
to alignment of the deuteron, we find strong dependence of FSI
effects on tensor polarization of the deuteron, which leads to a very
rich pattern of the angular and $p_m$ dependence of momentum
distributions and tensor analyzing power. In the tensor analyzing
power $A$, FSI effects turn out substantial at a relatively small
missing momentum $p_m \gsim (0.9-1.0) fm ^{-1}$ both in
transverse and longitudinal kinematics. In transverse kinematics,
for the longitudinal deuteron $\mu = 0$, we predict that FSI shifts
the dip of the momentum distribution from $p_m = 1.58 fm^{-1}$
to $p_m = 1.21 fm^{-1}$. For transverse
deuterons, $\mu = \pm 1$, FSI produces a shoulder-like structure
in $W_+(p_m, \theta = 90^o)$ as compared to a monotonous decrease
of $N_+ (p_m, \theta = 90^o)$ in PWIA. At large $p_m \gsim 2.5 fm
^{-1}$, the elastic rescattering peak in $W_+(p_m, \theta = 90^o)$
is stronger than in $W_o(p_m, \theta = 90^o)$, in agreement with the
expectation that the experimentally known sign of the quadrupole
deformation of the deuteron makes the probability of rescattering
of the struck proton on the spectator neutron larger for $\mu = \pm 1$.

A departure of the nuclear transparency ratio $T_{\mu} (p_m, \theta)
= \displaystyle {\frac {W_{\mu}(p_m, \theta)} {N_{\mu} (p_m, \theta)}}$
from unity is small, $T_{\mu} (p_m, \theta) \simeq 0.93$, only at
small $p_m \sim 0$. At larger $p_m$, in the region of strong
PWIA-FSI interference, one finds both $T_{\mu} (p_m, \theta) \ll 1$
and $T_{\mu} (p_m, \theta) \gg 1$. The antishadowing effect
$T_{\mu} (p_m, \theta) > 1$ is
particularly strong in transverse kinematics.

In PWIA, we find substantial sensitivity of the tensor analyzing
power to models of the deuteron wave function at large $p_m$.
FSI effects mask this sensitivity to a large extent. A unique
observable, which provides an FSI independent probe of the nodal
structure of the combination of S- and D-wave functions
$s(p_m) - \frac{1}{\sqrt{2}} d (p_m)$, is the forward-backward
asymmetry $A_{FB}^{(+)}$ for transverse deuterons $\mu = \pm 1$.
The asymmetry $A_{FB}^{(\mu)}$ is a pure PWIA-FSI interference
effect and has its origin in the nonvanishing real part of the
$np$ scattering amplitude. None the less, $A_{FB}^{(+)}$
is shown to be proportional to $s(p_m) - \frac{1}{\sqrt{2}} d(p_m)$,
and the node of $A_{FB}^{(+)} (p_m)$ is predicted to coincide
with the node of the PWIA amplitude
$s(p_m) - \frac{1}{\sqrt{2}} d(p_m)$.
Mention must be made of a potential sensitivity of FSI effects to
the spin-triplet $np$ total cross section in the state $\nu = 0$.

In conclusion, interpretation of the experimental data on
$^2 \vec H(e,e'p)$ scattering on tensor polarized deuterons at
missing momenta $p_m \gsim (0.9-1.0) fm^{-1}$requires an accurate
evaluation of FSI effects which depend strongly on the spin state
of the deuteron and on the scattering kinematics.

{\bf Acknowledgments:}
One of us (A.B.) acknowledges previous
discussions with S.Boffi on the general treatment of the
photon-deuteron interaction.
A.B. thanks J.Speth for the hospitality
at IKP. S.J. thanks S.Boffi for the hospitality at the
University of Pavia.
This Germany-Italy exchange program was supported in part
by the Vigoni Program of DAAD (Germany) and of the Conferenza
Permanente dei Rettori (Italy). This work was also supported by
the INTAS Grant No. 93-239 and by the Grant N9S000 from the
International Science Foundation.
\pagebreak\\


\pagebreak

{\bf \Large Figure captions:}

\begin{itemize}

\item[{\bf Fig.~1}]
    The full momentum distributions (solid line) $W_o$ (a),
    $W_+$ (b) including FSI and $N_o$ (c), $N_+$ (d)
    vs. $p_m$ at $\theta = 90^o$ and their decomposition in the
    S-wave (dotted line) and D-wave (dash-dotted line) contribution.
    The dashed line shows the sum of S- and D-wave contribution.

\item[{\bf Fig.~2}]
    The tensor analyzing power vs. the missing momentum $p_m$.
    Panel (a) shows $A^{PWIA}(p_m,\theta = 90^o)$ calculated
    using the Bonn wave function (dotted line) and
    calculated using the Paris wave function (dash-dotted line)
    and also $A(p_m, \theta = 90^o)$ calculated with the Bonn
    wave function (solid line) and with the Paris wave function
    (dashed line).
    Panel (b) shows $A^{PWIA}(p_m,\theta = 0^o) = A^{PWIA}
    (p_m, \theta = 180^o)$  (dotted line) and $A$ calculated
    including FSI for $\theta = 0^o$ (solid line) and
    $\theta = 180^o$ (long-dashed line).
\item[{\bf Fig.~3}]
    The angular dependence of the missing momentum distributions
    $N_o$ (dashed line), $N_+$ (dotted line),
    $W_o$ (solid line) and $W_+$ (dash-dotted line) for different
    values of the missing momentum $p_m$.

\item[{\bf Fig.~4}]
    The angular dependence of the tensor analyzing power in
    PWIA (dotted line) and including FSI (solid line) for
    different values of the missing momentum.

\item[{\bf Fig.~5}]
    The real parts of the matrix elements $s(\vec p_m)$ (solid line),
    $d_o (\vec p_m)$ (dotted line) and $d_2 (\vec p_m)$ (dashed line)
    for $\theta = 90^o$ including FSI, as defined in
    eqs. (\ref{ms}), (\ref{md}). Notice the change of the scale of
    $d_o(\vec p_m)$ at $p_m = 2.3 fm^{-1}$,
    and of $d_2 (\vec p_m)$ at $p_m = 2.7 fm^{-1}$.

\item[{\bf Fig.~6}]
    (a) The momentum distributions $W_+(p_m, \theta = 0^o)$ (solid
    line), $W_+(p_m, \theta = 180^o)$ (dashed line) and
    $W_o(p_m, \theta = 90^o)$ (dash-dotted line) including FSI
    and $N_+(p_m, \theta = 0^o) = N_+(p_m, \theta = 180^o)$
    (dotted line).
    (b) The momentum distributions $W_o(p_m, \theta = 0^o)$ (solid
    line), $W_o(p_m, \theta = 180^o)$ (dashed line) and
    $N_o(p_m, \theta = 0^o) = N_o(p_m, \theta = 180^o)$
    (dotted line).

\item[{\bf Fig.~7}]
    The momentum distributions calculated with the Bonn and Paris
    wave functions in PWIA (b) and including FSI (a) at
    $\theta = 90^o$. The results calculated with the Bonn wave
    function for $W_o$ and $N_o$ are shown by solid lines,
    $W_+$ and $N_+$ are shown by dash-dotted lines,
    the results calculated with the Paris
    wave function for $W_o$ and $N_o$ are shown by the
    dashed lines and for $W_+$ and $N_+$ are shown by
    dotted lines.

\item[{\bf Fig.~8}]
    (a) The forward-backward asymmetry $A_{FB}$ vs. $p_m$ for
    $W_+$ calculated with the Bonn wave function (solid line)
    and the Paris wave function (dashed line).
    (b) The forward-backward asymmetry $A_{FB}$ vs. $p_m$ for $W_+$
    (solid line) and $W_o$ (dotted line) calculated with the
    Bonn wave function.

\end{itemize}
\end{document}